A Possible Explanation for the Blue Spectral Slope Observed on B-type Asteroids

M.J. Loeffler[1,2], and  B.S. Prince[3]

[1]Department of Astronomy and Planetary Science, Northern Arizona University, Flagstaff, AZ, 86011

[2]Center for Materials Interfaces in Research and Applications, Northern Arizona University, Flagstaff, AZ, 86011

[3]Department of Applied Physics and Materials Science, Northern Arizona University, Flagstaff, AZ, 86011

Abstract

In an effort to better understand the role dark material plays in the reflectance spectrum of carbonaceous asteroids, we performed laboratory studies focusing on quantifying how the addition of relevant dark material (graphite, magnetite and troilite) can alter the ultraviolet-visible and near-infrared spectrum of a neutral silicate mineral. We find that addition of graphite, magnetite and troilite all darken the reflectance spectrum of our forsterite samples and cause the spectral slope to decrease (become blue). These spectral changes can be caused by both nm- and μm-sized grains. In the ultraviolet-visible region, we find that graphite is most efficient at altering the spectral slope, while in the near-infrared, magnetite is the most efficient. At all wavelengths studied, graphite is the most efficient at darkening our sample spectrum. However, the observation that troilite also alters the slope and albedo of our samples suggests that the spectral changes caused by magnetite and graphite may not be unique. In addition, we find that the spectral slopes in our mixtures compare generally well to what has been observed on Bennu suggesting that a significant portion of fine-grained dark material, including sulfides, present in the regolith can cause the observed negative (blue) slope found on B-type asteroids.

1. Introduction

For decades, remote sensing reflectance spectroscopy has been an enormously valuable technique for classifying and characterizing the surface of airless bodies (Gaffey et al. 1993, Bus and Binzel 2002, Demeo et al. 2009). Due to a number of factors, the most widely utilized spectral region has been the near-infrared ($0.7 - 2.5$ μm), and within this spectral region, a key characteristic that is used for analysis is the spectral slope of an object. Drawing on our understanding of the lunar surface and of relevant laboratory experiments (Hapke 2001, Sasaki et al. 2001, Brunetto and Strazzulla 2005, Loeffler et al. 2009), a positively (red) sloped spectra in the near-infrared has been often tied to the presence of chemically reduced (metallic) iron, which has been hypothesized to be formed via space weathering processes, such as solar wind and micrometeorite impacts. Similar arguments have been made to explain the spectra of many asteroids, although it seems that the effects of space weathering on these surfaces may differ somewhat from "lunar-style" weathering (Chapman 2004, Gaffey 2010, Brunetto et al. 2015).

Interestingly, many carbonaceous asteroids, namely B-types (Demeo et al. 2009, Clark et al. 2010), have a negative (blue) spectral slope in the visible and near-infrared region. Early studies using remote sensing data from the Sloan Digital Sky Survey (SDSS) dataset suggested that the blue spectral slope is correlated with the age of the asteroid (Nesvorny et al. 2005). Although the trend is different than what has been previously attributed to space weathering (blue vs. red spectral slopes), a correlation between slope and age would be consistent with a time dependent process such as space weathering, as has been suggested by a number of studies (Moroz et al. 2004, Vernazza et al. 2013, Matsuoka et al. 2015, Gillis-Davis et al. 2017, Kaluna et al. 2017, Lantz et al. 2017, Thompson et al. 2020, Trang et al. 2021). Interestingly, Thomas et al. (2021) have recently used the SDSS dataset to show that even within the same asteroid family, the spectral slope can vary significantly. This suggests that other factors, in addition to space weathering, may make an important contribution to the spectral slope on carbonaceous asteroids. One possible factor, first examined by Johnson and Fanale (1973) and revisited in more detail by others (Miyamoto et al. 1982, Clark 1983, Milliken and Mustard 2007, Clark et al. 2011, Cloutis et al. 2011a, Cloutis et al. 2011b, Cloutis et al. 2012), is that fine-grained opaque materials (e.g. magnetite and some forms of carbon) mixed with silicates could cause a blue spectral slope in the near-infrared region. It has also been shown that the spectral slope of some carbonaceous chondrites can depend on grain size, as some larger grained carbonaceous chondrites with blue or neutral slopes in the visible and near-infrared become red after they are crushed to a smaller grain size (Johnson and Fanale 1973, Cloutis et al. 2011a, Beck et al. 2021). Recently, it has been postulated that this grain size effect may responsible for observed slope differences among C-type asteroids (Beck and Poch 2021).

Given the recent interest in B-type asteroid 101955 Bennu, which also has a blue spectral slope in the near-infrared (Clark et al. 2011, Hamilton et al. 2019, Simon et al. 2020b), understanding the possible nature and origin of this identifiable characteristic is of even greater interest. Drawing on the likelihood that the dark material present in the regolith could be key to producing a blue spectral slope, we performed laboratory studies aimed at understanding if another opaque phase, besides carbon and magnetite, could also cause an asteroid surface to appear blue. We also investigated how efficient each of these opaques would be in reference to one another and how these spectral changes depend on the grain size of the dark material. For simplicity, we used synthetic forsterite (Fo$_{99+}$) as our starting sample material and made binary mixtures containing specific amounts of carbon (graphite), Fe$_3$O$_4$ (magnetite), or FeS (troilite), while monitoring the spectral reflectance of our sample mixtures between 0.25 and 2.5 μm. Resulting spectra are compared with one another and previous works and are discussed in the context of some recent published spectra of Bennu.

## 2. Materials and Methods

### 2.1 Samples and sample preparation

In this study, we used synthetic forsterite (Fo$_{99+}$), graphite (C), magnetite (Fe$_3$O$_4$) and troilite (FeS). We purchased Fo$_{99+}$ from Reade Advanced Materials, and it has estimated impurities (in wt%) of Fe$_2$O$_3$ (0.02), Al$_2$O$_3$ (0.03), TiO$_2$ (0.09), Na$_2$O (0.11), CaO (0.22), and K$_2$O (0.12). We purchased 1-cm graphite rods from Alfa Aesar (99% purity), 400 – 500 nm x 40 nm graphite grains from ACS Material (purity > 99%) and 3 μm, 10 μm, and 25 μm spherical graphite grains from US Research Nanomaterials (purity 99.95%). We purchased magnetite (99.99% purity) in the form of a powder and synthetic troilite rods[1] from Sigma Aldrich. Although no purity estimates were given for troilite, our prior analysis has shown the surface composition of these freshly crushed rods are consistent with what is expected (Loeffler et al. 2008).

To prepare our samples, we used agate mortar and pestle to grind our starting material (if needed) and dry-sieved it to the appropriate grain size. The dark material (troilite, graphite, magnetite) was sieved to grain sizes <45 μm, while the Fo$_{99+}$ powder was sieved to either 45 – 125 μm or 500 – 600 μm depending on the experiment. We did not sieve or alter the graphite that was purchased as nm- or μm-sized grains. As we have seen previously that prolonged exposure to atmosphere will cause the surface sulfur in troilite to oxidize (Loeffler et al. 2008), all reflectance measurements involving troilite were done immediately following crushing and sieving the synthetic rods to the desired grain size (<45 μm).

To properly quantify the composition of our mixtures, we used a high-precision analytical balance (JJ224BC) with a specified resolution of 0.1 mg. To determine the initial mass of the sample, we weighed a glass vial, added Fo$_{99+}$ to the vial, weighed the vial again and then took the mass difference. Starting samples typically had a mass of ~200 mg. To prepare for reflectance analysis (see below), sample powder was loosely poured into a 10-mm aluminum ring, and the surface of the sample was carefully scraped to ensure that any compression of the sample material was minimal. After reflectance analysis, the powder was poured back into the glass vial, reweighed, and a known amount of dark material was added. As the dark material was typically added in relatively small amounts (1 – 10 mg), we made a small cup out of aluminum foil to separately determine the mass of the dark material.  Next, we poured the known amount of dark material into the glass vial, closed the top of the vial, and shook it vigorously for 1 min.  We followed this by gently stirring the mixture with a spatula, which we found was the most effective method for homogenizing our mixture. Once we achieved sufficient mixing, the resulting sample mixture was removed and prepared for reflectance analysis, as described above. To be consistent with previous studies (e.g., Clark et al. 2008, Cloutis et al. 2011a), we are labeling the sample mixture in terms of the wt% of dark material present.  Of course, given that the mass of one mole of graphite (12 g) is significantly lower than one mole of troilite (88 g) or magnetite (232 g), for the same wt% there are about seven (or nineteen) times more carbon atoms than there are troilite (or magnetite) molecules.

---

[1]Our two prior publications on FeS (Loeffler et al. 2008, Prince et al. 2020) stated that the FeS was purchased from Alfa Aesar.  However, those samples, as well as the ones used here, were purchased from Sigma Aldrich (product #12363).

We note that visual inspection of our mixtures revealed that the samples composed primarily of 45 – 125 μm $Fo_{99+}$ appeared to be uniform. This observation is supported by the consistency of the reflectance spectrum taken of our mixtures at the end of each experiment. Specifically, we measured the final reflectance spectrum of these mixtures five times. Between each measurement, we returned the mixture to our sample vial, mixed it as described above, and extracted approximately half of the sample for our subsequent measurement. Across the entire wavelength range these spectra deviated no more than 5% from the average value taken in those measurements. While the smaller-grained mixtures appeared to mix well, the ones composed of 500 – 600 μm $Fo_{99+}$ did not.  For instance, for the 500 – 600 μm $Fo_{99+}$ samples, small amounts of dark material (~0.5 wt%) appeared to preferentially stick to the surface of the larger grains, while addition of more material caused clear separation of the phases that could not be rectified by our mixing. Thus, we consider the wt% reported for the 500 – 600 μm $Fo_{99+}$ samples to be a qualitative estimate of our mixture composition.

## 2.2 Reflectance Analysis

We analyzed our samples under ambient conditions with ultraviolet-visible (uv-vis) and near-infrared reflectance (NIR) spectroscopy. The uv-vis measurements were made with an Avantes 2048XL fiber optic spectrometer over the wavelength range of 0.25 – 0.9 μm, at a resolution of 0.0015 μm. The dwell time of the detector for a single scan was 0.5 sec, and each measurement is an average of 40 scans. As is standard practice with these fiber optic spectrometers, both the source and detector fiber optics were mounted to collimating lenses, which served to both collimate the beam as it left the source and to focus the reflected signal back onto the fiber optic connected to the detector; the spot size on the sample was measured to be ~3.7 mm in diameter. The NIR measurements were made with a Thermo-Nicolet iS50 Fourier Transform Infrared Spectrometer (FTIR) over the range of 13,400-4000 $cm^{-1}$ (0.75 - 2.5 μm) at a spectral resolution set to 16 $cm^{-1}$, and each spectrum is an average of 400 scans. In those cases, a halogen light source was focused onto our sample down to a spot size of ~2.7 mm, while the reflected light was focused onto an InGaS detector. In both cases, light was aimed vertically downward directly at the sample's surface (0° incidence) and reflected light was collected at 30°. The resulting reflectance spectra are given by

$$R_{uv-} = \frac{I_{sample} - I_{dark}}{I_{reference} - I_{dark}} \qquad\qquad R_{NIR} = \frac{I_{sample}}{I_{reference}} \qquad (1)$$

where $I_{sample}$ is the reflected intensity from the sample, $I_{reference}$ is the reflected intensity from our reference, and $I_{dark}$ is detector signal when the light source is blocked. In both cases, the reference sample was loose PTFE (Teflon) powder (Avocado Research Chemicals). The PTFE sample has been calibrated against an absolute reflectance standard (Spectralon; Lapshere certified reflectance standard), allowing us to report the absolute reflectance of our sample.

Using the derived reflectance spectra from (1), we quantify the spectral changes induced by adding the dark material to our $Fo_{99+}$ sample by analyzing the spectral slope and overall albedo of the reflectance spectrum. The spectral slope reported in this study has been calculated by evaluating the derivative of the spectrum and taking the average value over two regions representative of the uv-vis ($0.26 - 0.5$ μm) and the near-infrared ($0.8 - 2.4$ μm). As in our previous study (Prince et al. 2020), the overall albedo has been calculated by comparing the initial reflectance of the sample ($R_0$) with the reflectance of the sample mixture (R) at the wavelength of interest. We refer to this as the normalized reflected intensity, which is defined as $R/R_0 - 1$.

Prior to being able to use the absolute reflectance spectra, spectral slope and normalized reflected intensity to compare how efficiently the different dark materials alter the $Fo_{99+}$ sample, we verified the reproducibility of the observed spectra and resulting trends. For our initial spectra, we found that the absolute reflectance was reproducible to ~5%, while the spectral slope showed even less variation from sample to sample. For our given sample mixtures, we found that the reported trends were reproducible, although there was some variation in the reflectance spectra and derived spectral parameters, which is expected. More specifically, we found that a sample mixture with > 1 wt% of dark material had reflectance spectra and derived spectral parameters that were reproducible to within about 15%. In cases where the mixture contained ≤ 1 wt% of dark material, the reflectance spectra and derived spectral parameters were somewhat more variable, yielding differences as high as 30% in some cases. We attribute the larger deviation at low wt% to be mainly a result of the uncertainty in the mass of the added dark material (estimated to be ± 0.5 mg), which at these early data points is a significant amount of the added material (~1 mg).

*2.3 Microscope Analysis*

In order to better constrict the grain size of our different dark materials, we performed scanning electron microscopy using a Zeiss Supra 40VP field emission scanning electron microscope (SEM). For the sieved materials, we made rudimentary estimates of the average grain size by counting the number of grains on a given line and dividing by the length of the line. We did this for multiple regions of these sample to obtain an average grain size.

3. Results

*3.1 Initial Characterization of the Single Components*

*3.1.1 Reflectance Analysis*

Figure 1 shows the absolute reflectance spectra of the sieved $Fo_{99+}$ (45-125 µm), graphite (<45 µm), magnetite (<45 µm) and troilite (<45 µm) samples used in this study. The spectrum of $Fo_{99+}$ is largely featureless and bright in the visible and near-infrared region, but has an absorption band at lower wavelengths, which we previously attributed to a low amount of iron impurities in the sample that are likely in the form of either $Fe^{2+}$ or $Fe^{3+}$ (Loeffler et al. 2016).  The spectrum of graphite is relatively featureless, dark and positively sloped in the near-infrared region. The sharp rise in reflectance below 0.5 µm is due to a Fresnel peak near 0.25 µm caused by a strong absorption resulting from the π-π* electronic transition (McCartney and Ergun 1967, Applin et al. 2018). The overall spectrum of magnetite is red and dark and contains a broad absorption band near 1 µm, as well as a weak feature centered near 0.55 µm, due to $Fe^{2+}$ and $Fe^{3+}$, respectively (Cloutis et al. 2011a).  The spectrum of synthetic troilite is consistent to what has been measured for meteoritic troilite (Britt et al. 1992, Cloutis et al. 2011b).  Specifically, it is relatively featureless, dark and positively sloped in the near-infrared region, but contains an absorption band centered near 0.35 µm, which has been attributed to a combination of Fe-S charge transfer and crystal field transitions of $Fe^{2+}$ (Marfunin et al. 1968, Vaughan and Craig 1978).

Figure 2 shows the absolute reflectance spectra of the graphite grains with well-constrained grain sizes that we used in this study. The spectra of these powders are generally similar to the spectrum of the <45 µm graphite powder shown in Figure 1, although the nm-sized graphite powder is darker and has a flatter spectral slope above 1.5 µm compared to the other powders.

*3.1.2 SEM Images of the Dark Material*

Figure 3 shows SEM images representative of the <45 µm graphite, magnetite and troilite sieved material characterized in Figure 1. Generally, these images show that the grains mostly angular and, in some cases, have some small particles adhered to the surface of some of the larger grains.  In addition, the graphite and troilite grains appear to be smoother than the magnetite grains, although some of the graphite grains appear to be agglomerates. To make a rough estimate of the average grain size of the bulk of each sample, we assumed agglomerates were a single grain and ignored the small particles (< 1 µm) that were adhered to the larger grains.  Using the method described in *Section 2.3* we estimate that the average grain size of these materials are: 12 µm (graphite), 10 µm (magnetite) and 7 µm (troilite).

Figure 4 shows SEM images representative of the well-constrained µm-sized graphite grains that we used in this study. In contrast to the <45 µm sieved material, these grains are smoother and more round, although not completely spherical. Furthermore, although the majority of the particles observed in each sample are consistent with the manufacturer specifications, all samples have some size distribution associated with them. Figure 5 shows SEM images representative of the well-constrained 400-500 nm x 50 nm graphite grains.  Although the imaging of these grains is more difficult due to their small size, their size appears to be consistent with the manufacturer specifications (Figure 5 bottom), although in some cases they appear to be agglomerates (Figure 5 top).

## 3.2. Reflectance Analysis of Mixtures

### 3.2.1 The Effect of Adding Different Dark Materials

Figure 6 shows the absolute reflectance spectra of $Fo_{99+}$ (45 – 125 μm) samples with varying amounts of sieved <45 μm graphite, magnetite, or troilite mixed into the matrix (0 – 15 wt%). Generally, the range of wt% was chosen to represent what would be expected for an asteroid regolith based on measurements of carbonaceous chondrites, which estimate the carbon, magnetite and sulfur components to be as much as 6 wt% (Pearson et al. 2006), 11 wt% (Hyman and Rowe 1983) and 8 wt% (Nittler et al. 2004), respectively. Figure 6a shows the effect of adding graphite to the $Fo_{99+}$ sample. Not surprisingly, addition of the graphite to the $Fo_{99+}$ samples causes the spectral reflectance to decrease. However, instead of causing an increase in spectral slope (reddening), which might be expected based on the examination of the pure graphite spectrum, the addition of graphite causes the spectral slope of the sample to decrease (become blue) in the visible and near-infrared region, in agreement with previous work (e.g. Cloutis et al. (2011a)). A nearly identical effect is also observed when we replaced graphite with magnetite (Figure 6b), although qualitatively it appears that graphite is slightly more efficient in altering the spectrum of our silicate sample. As with the other two minerals, adding troilite to $Fo_{99+}$ causes the same trend of darkening and bluing (Figure 6c). However, qualitatively troilite appears to be the least efficient of these three dark minerals in causing spectral changes in our silicate sample, which may not be unexpected given its albedo in this region is about a factor of two higher than graphite and magnetite. Finally, we investigated whether this trend would also be observed if the silicate material in the regolith contained larger grains by adding graphite to a 500 – 600 μm $Fo_{99+}$ sample (Figure 6a). Although achieving homogeneous mixing was problematic in these samples (see *Section 2.1*), we observe the same general trend of darkening and bluing of the reflectance spectrum with the addition of graphite.

To more quantitatively compare the spectral changes induced by addition of each dark material, we plot the average spectral slope over a portion of the uv-vis (0.26 – 0.5 μm) and near-infrared (0.8 – 2.4 μm), as well as the normalized reflected intensity (see *Section 2.2*) at 0.35, 1.0 and 2.0 μm, in Figure 7 and 8. The addition of any of our dark material causes the spectral slope to decrease in both spectral regions. In the near-infrared, the slope goes from being approximately neutral to blue, and the majority of the change occurs after adding only a small amount of dark material (~1 wt%). Furthermore, adding magnetite causes the spectral slope to be ~30% more blue than graphite and troilite in this spectral region. In the ultraviolet-visible, the spectral slope becomes significantly less red as dark material is added, and the decrease in spectral slope is more gradual than is observed in the near-infrared. Unlike in the near-infrared, graphite appears to be the most efficient in altering the spectral slope in the uv-vis, as at the highest concentrations studied, the spectral slope in the graphite mixtures is ~2.5 times lower than in the two other mixtures. As with the spectral slope, all dark materials cause the normalized reflected intensity to decrease (i.e. the sample darkens). Although the darkening effects are somewhat similar between the three dark materials, graphite appears to be the most efficient of the three samples, causing darkening between ~50 and 70% depending on the wavelength.

*3.2.2 The Effect of Grain Size*

Figure 9 shows the absolute reflectance spectra of $Fo_{99+}$ (45 – 125 μm) samples with varying amounts of graphite grains added to the matrix (0 – 10 wt%) for grain sizes of 400 – 500 nm x 40 nm, 3 μm, 10 μm, and 25 μm. As with our other mixtures, all spectra darken and show a decrease in slope as we add graphite to the sample. Qualitatively, both of these changes appear to happen most quickly in the mixture containing nm-sized graphite (Figure 9 top), although the spectral slope in the near-infrared also appears to flatten out (increase) slightly at higher graphite concentrations in those mixtures. The three mixtures containing μm-sized grains are relatively consistent with one another, yet the two smaller grain-sizes (3 and 10 μm) appear to more efficiently alter the $Fo_{99+}$ reflectance spectrum than do the 25-μm grains.

To quantitatively compare the spectral changes observed in Figure 9, we plot the average spectral slope and normalized reflected intensity as a function of graphite wt% in Figure 10 and Figure 11 for the different grain sizes of graphite studied in this work. These more quantitative metrics reinforce our initial observations that the changes in spectral slope and normalized reflected intensity are similar to what we observed in our mixtures containing <45 μm grains and that the mixture containing nm-sized graphite is more efficient in causing spectral changes than the mixtures containing the μm-sized grains. More specifically, after adding ~2 wt% of graphite the mixtures containing 3, 10 and <45 μm grains have a spectral slope in the ultraviolet-visible that is within 10% of one another, is about 30% lower than mixture containing 25 μm grains but is about three times higher than the slope in the mixture containing the 400 - 500 nm grains. In addition, the maximum slope observed in the near-infrared for the nm-sized graphite mixtures is ~60% lower than in the mixtures containing μm-sized graphite, although with increasing graphite concentration the trend in the nm-sized mixtures is unique: the slope increases and ends up being comparable to the slope measured for the μm-sized mixtures. Finally, the nm-sized graphite causes darkening of the forsterite between ~80 and 90 %, the 3, 10 and <45 μm grains cause darkening between ~50 and 70 %, and the 25 μm grains cause darkening between 35 – 50 % depending on the wavelength.

4. Discussion

*4.1 The Origin of the Blue Spectral Slope*

Figures 6 and 9 show that the adding any of our three chosen dark materials to our bright neutral starting material results in a mixture that is darker and has a blue spectral slope in the near-infrared region. While the trend of the former is not surprising, the latter trend shows that the addition of this dark material is more complicated than treating the resulting spectrum as a linear combination of the two components, otherwise all the resulting spectra would have a red spectral slope in the near-infrared region. Previous laboratory work, focused on understanding the spectral slope observed on some of the Saturnian satellites, showed that when very small (200 nm) grains of dark material (lampblack) are mixed with ice, the spectral slope in the near-infrared turned blue, and this change was attributed to Rayleigh scattering (Clark et al. 2008). In addition, Clark et al. (2008) also observed that there was a concentration where the blue spectral slope reached a maximum value (~2 wt% of lampblack), which, given the very different nature of our two starting materials (ice vs. forsterite), is consistent with what we observed in our experiments with nm-sized graphite. Another recent study, aimed at understanding the blue spectral slope observed on regions of Ceres, suggests that the formation of porosity in salty material through sublimation can also cause the spectral slope in this region to appear blue (Schroder et al. 2021). Again, the porosity in those experiments appears to be on a small enough scale that it can also be explained by Rayleigh scattering.

The potential role of Rayleigh scattering cannot be completely ruled out in our experiments using the sieved dark materials, as only the upper end of the grain size is well-constrained (< 45 µm). Furthermore, SEM images of those samples suggest that there could be some particles present that are significantly smaller than the estimated average grain size. However, the experiments on graphite grains with a well-constrained grain size support that the spectral bluing observed in our experiments is not due to Rayleigh scattering, as only the 400 – 500 nm x 50 nm grains would meet the Rayleigh criteria (Hapke 2001, Brown 2014), and this would only be in cases when the grains are oriented such that they appear to be the size of the sample thickness (50 nm). Instead, in our experiments it seems that adding any of these µm-sized dark, slightly red materials is sufficient to cause the spectral slope of our starting neutral material to turn blue. Furthermore, the observation that the magnitude of the alteration in spectral slope is very similar in all of our mixtures containing µm-sized graphite suggests that, in this grain size range, the optical properties and the amount (wt%) of dark material in the mixture is more important than the grain size of the dark material in causing the observed spectral changes. Finally, the similarity of the spectral changes in our <45 µm graphite experiments compared with the 3, 10 and 25 µm experiments also supports that Rayleigh scattering is not what is driving the spectral changes in any of the experiments involving <45 µm sieved grains, even if there may be very small particles inevitably mixed with these samples. This similarity also suggests that the shape of dark material's grains (angular/rough vs. rounded/smooth) does not have a strong effect on the reflectance spectra of our mixtures.

## 4.2 Comparison to Previous Experiments

As noted earlier, the role that magnetite and graphite may play in causing the spectra of many carbonaceous chondrites and silicate mixtures to appear blue has been shown previously (Johnson and Fanale 1973, Miyamoto et al. 1982, Clark 1983, Milliken and Mustard 2007, Clark et al. 2011, Cloutis et al. 2011a, Cloutis et al. 2011b, Cloutis et al. 2012). Our results on these two materials are generally consistent with those studies, although we show that which material is most efficient at altering the spectral slope will depend on the wavelength region of interest. This consistency of results, regardless of what type of silicate (olivines or phyllosilicates) is used as a matrix mineral, supports that these spectral alterations could be observed on any airless body that contains these dark materials. In addition, the observation that troilite causes similar changes to the spectral slope and overall albedo illustrates that these observed trends are not uniquely tied to magnetite or graphite. In fact, given that troilite has been found in abundances up to ~8 wt% in carbonaceous chrondrites (Nittler et al. 2004), we speculate it could also play a very important role in causing the observed blue spectral slope on B-type asteroids. Further generalizing this observation to draw conclusions about any dark material present in the asteroid regolith is difficult, as some forms of carbon (lampblack) have shown to cause inconsistent changes to these spectral properties (Johnson and Fanale 1973, Clark 1983, Milliken and Mustard 2007, Cloutis et al. 2011a) and other relevant dark materials, such as nanophase iron and organics, may cause spectral reddening. In fact, this interplay between these "reddening" and "bluing" agents may explain the variation in spectral slope that has been observed among carbonaceous asteroids, although other factors could also be contributing, such as space weathering or a variation in grain size (see *Section 1*).

*4.3 Comparison to Remote Sensing Observations of Bennu*

Although our sample mixture is relatively simple, it is of interest to determine whether the spectral slope in our laboratory mixtures is sufficient to mimic the blue spectral slope observed on B-type asteroids. In Figure 12, we compare our results to an unresolved globally averaged spectrum of asteroid Bennu after normalizing both data sets to unity at 0.55 μm. We find that while the blue spectral slope is fairly representative of what is observed for Bennu, we can only match certain portions of Bennu's spectrum with our laboratory mixtures. For instance, Bennu's spectrum is well fit above 2.0 μm in samples containing 2 - 3 wt% of magnetite or graphite or in ones containing very high, and possibly unrealistic, amounts of troilite (15 wt%).  At shorter wavelengths (< 1 μm), Bennu's spectrum is better fit with samples that contained 8 - 9 wt% of magnetite or graphite but could not be fit with any of our troilite mixtures.

Even though Bennu's blue spectral slope seems to be generally reproduced by our laboratory data and portions of the spectrum can be fit fairly well, our mixtures are likely too simple to expect to be able to match Bennu's entire near-infrared spectrum. For instance, although recent studies support that Bennu is rich in magnetite (Simon et al. 2020a, Hamilton et al. 2021), which suggests it is largely responsible for the observed blue spectral slope, other dark material will be present on Bennu's surface. Possibilities include organics (Simon et al. 2020a), which may act to offset some of the spectral bluing caused by magnetite, or even graphite or troilite (Simon et al. 2020b), which we show would contribute to the spectral bluing.  Regardless, even with a combination of differing amounts of dark material in our silicate sample, the spectrum in this region would not likely be any better fit to Bennu, because of the concavity that is centered near 1.0 μm. The exact origin of this concavity is unclear, but most recent measurements of Bennu's surface taken with higher spatial resolution (Simon et al. 2020a) open up the possibility that this concavity is due to multiple absorptions that are simply unresolved in the globally averaged spectrum of Bennu. The multiple absorptions have been speculated to be due to a number of minerals, including magnetite and phyllosilicates, among others (Simon et al. 2020a). In fact, an earlier study noted the possibility that this broad feature may be an absorption band but the signal-to-noise ratio was not sufficient to be more conclusive (Clark et al. 2011). Thus, in the future we hope to examine more complex mixtures, which are not only representative of the spectral slope on Bennu, but also can reproduce the putative absorption features that have recently been identified on the surface of Bennu.

5. Conclusions

Here we have studied the spectral effects caused by adding relevant dark material to a neutral silicate mineral in an effort to better quantify the role dark material plays in the reflectance spectrum of carbonaceous asteroids. We find that addition of graphite, magnetite, and troilite all darken the reflectance spectrum of our $Fo_{99+}$ samples and cause the spectral slopes to decrease (become blue). Our results are generally consistent with what has been found for previous studies focusing on graphite and magnetite. However, the observation that troilite, an important component in carbonaceous asteroids, can also cause similar spectral alteration suggests that these spectral changes are not uniquely tied to any specific mineral. In addition, we also find that spectral bluing can be caused by μm-sized graphite grains that are much larger than would be required to meet the criteria for Rayleigh scattering. Furthermore, the finding that the magnitude of the alteration in spectral slope is very similar in all of our mixtures containing μm-sized graphite suggests that, in this grain size range, the optical properties of the material and the amount (wt%) in the mixture is more important than the grain size or shape of the dark material.

The similarity of the spectral slopes in our mixtures to that observed on Bennu suggests that any fine-grained dark material present in the regolith can cause the observed negative (blue) slope found in the visible and near-infrared region of B-type asteroids. However, more accurately fitting the entire near-infrared region of asteroid Bennu or other B-type asteroids will, not unexpectedly, take a more complex sample mixture than what has been presented here. Future studies will focus on this aspect, while also considering the potential role that other processes, such as space weathering or variations in grain size, may also play in altering these sample mixtures.

Acknowledgements

BSP was supported by Northern Arizona University's Hooper Undergraduate Research Award. MJL and BSP thank MS Thompson for the use of the magnetite powder and VE Hamilton for the normalized spectrum of Bennu. Data from this publication can be found in Northern Arizona University's long-term repository (https://openknowledge.nau.edu/5614/).

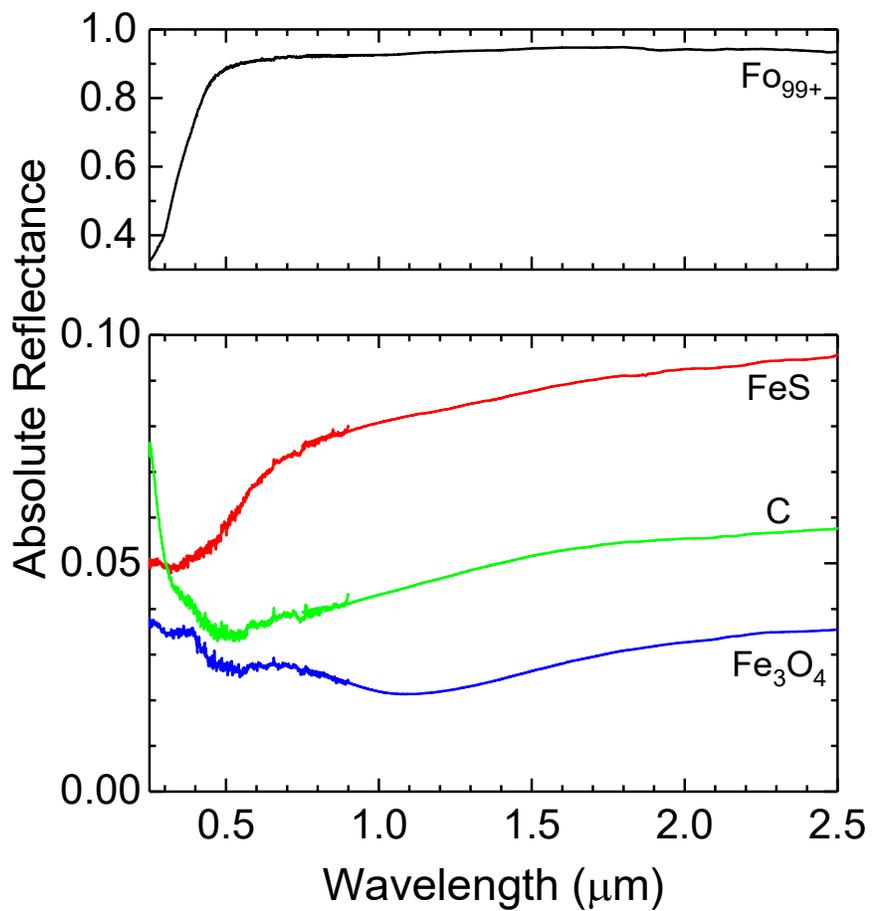

Figure 1. Absolute reflectance of a loose powder samples used in this study. Top: Fo$_{99+}$ (45 – 125 µm).  Bottom: FeS (< 45 µm), graphite (< 45 µm), and Fe$_3$O$_4$ (< 45 µm).

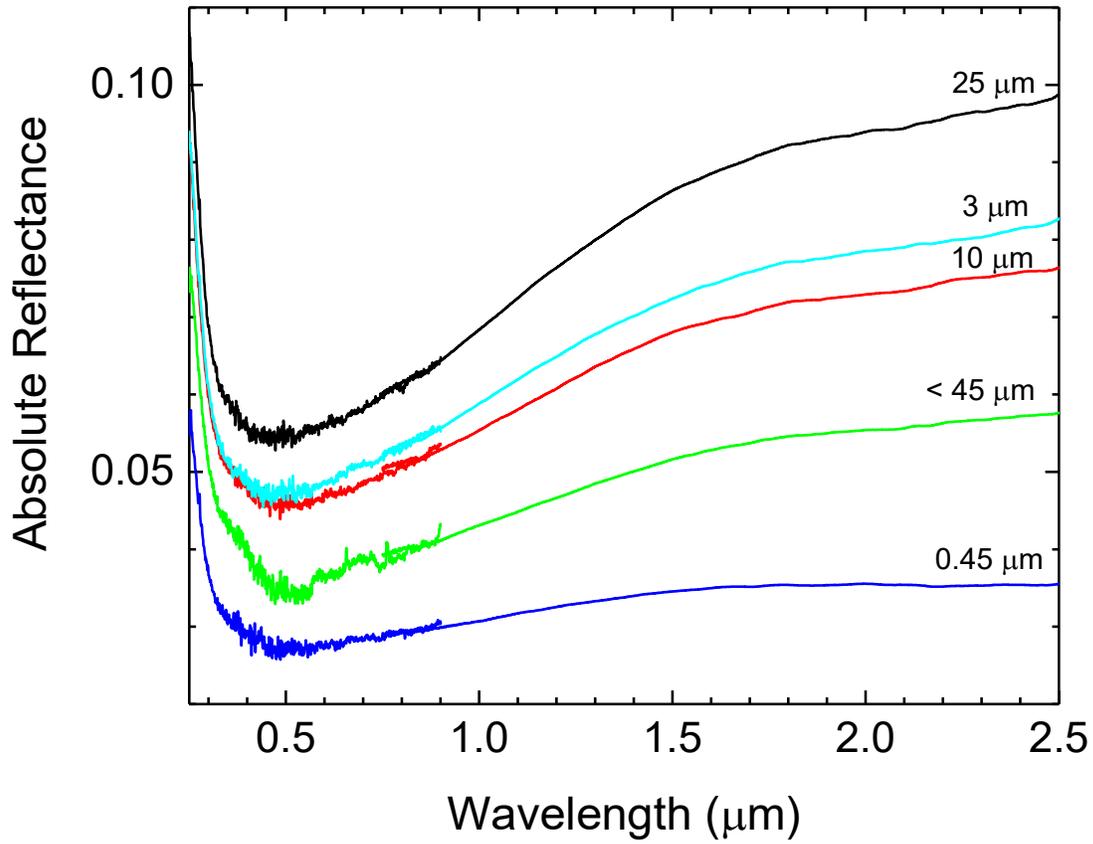

Figure 2. Absolute reflectance of the loose graphite powder samples used in this study. The corresponding grain size of each the sample measured is given in the figure. We note that the 400 – 500 nm x 50 nm have been labeled as 0.45 µm in the figure.

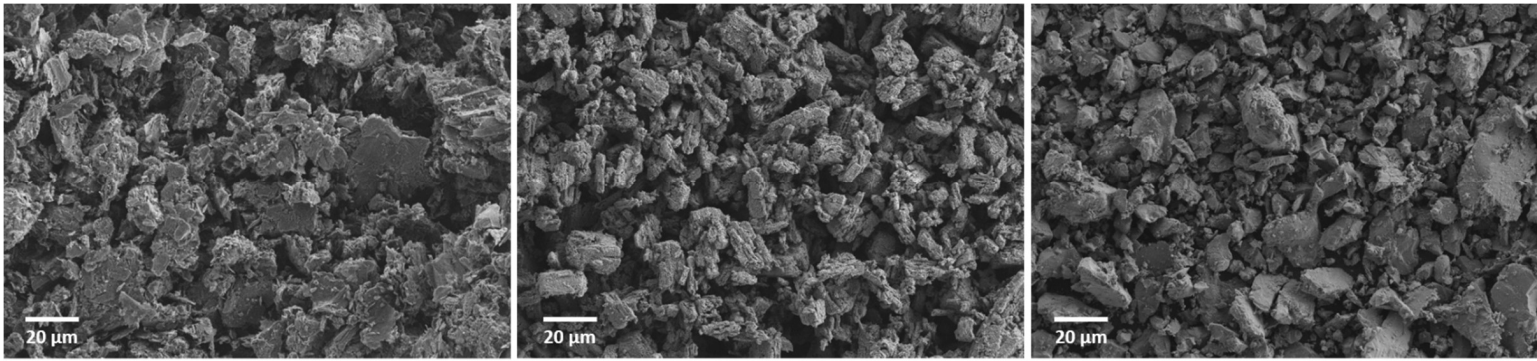

Figure 3.  SEM images of the <45 μm sieved dark material used in this study: graphite (left), magnetite (middle) and iron sulfide (right).

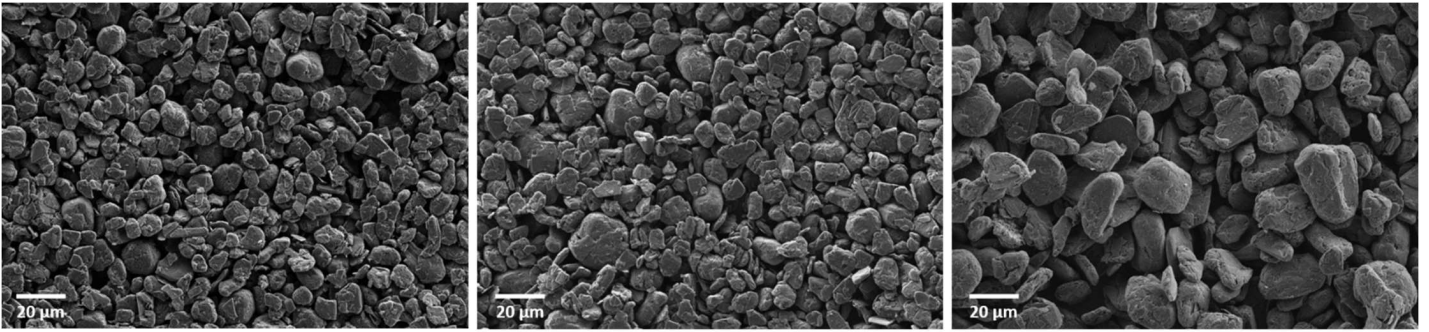

Figure 4.  SEM images of graphite with a well-defined grain size used in this study: 3-μm (left), 10-μm (middle) and 25 μm (right).

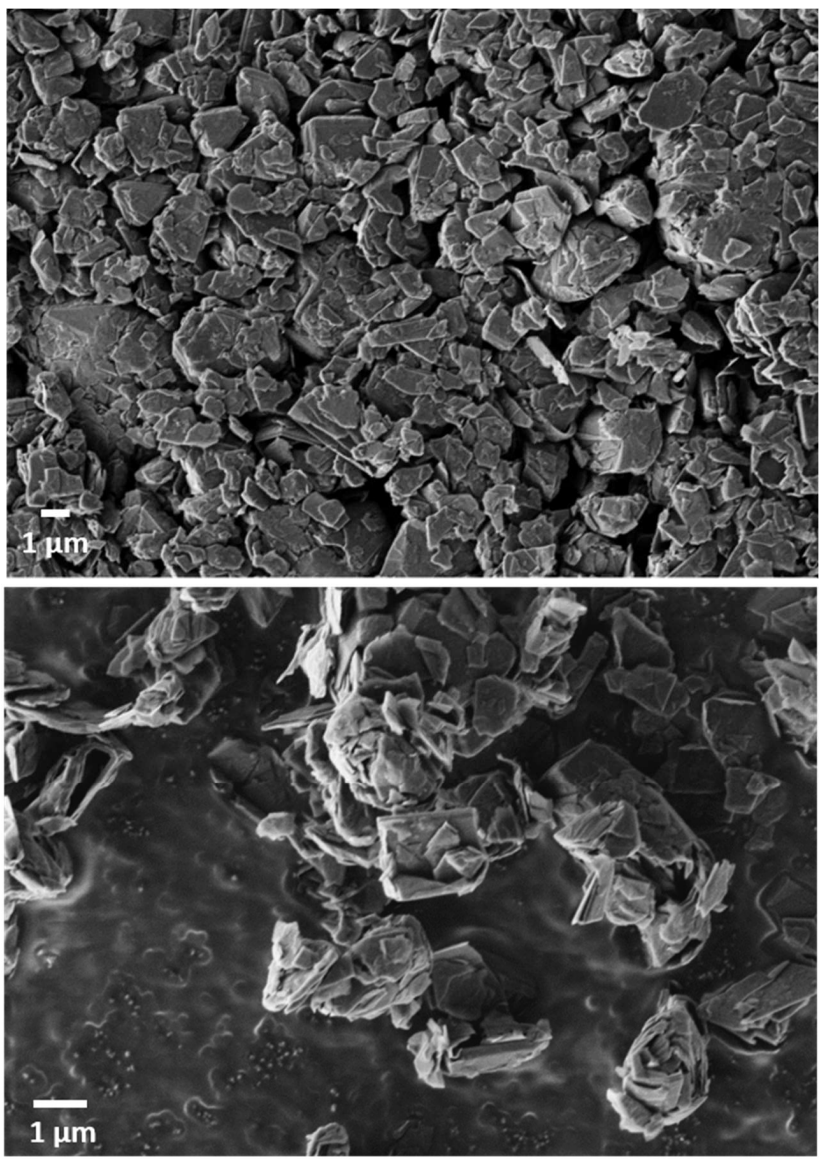

Figure 5.  SEM images of the 400 – 500 x 50 nm graphite grains used in this study. The smooth regions in the bottom image is from the underlying carbon tape.

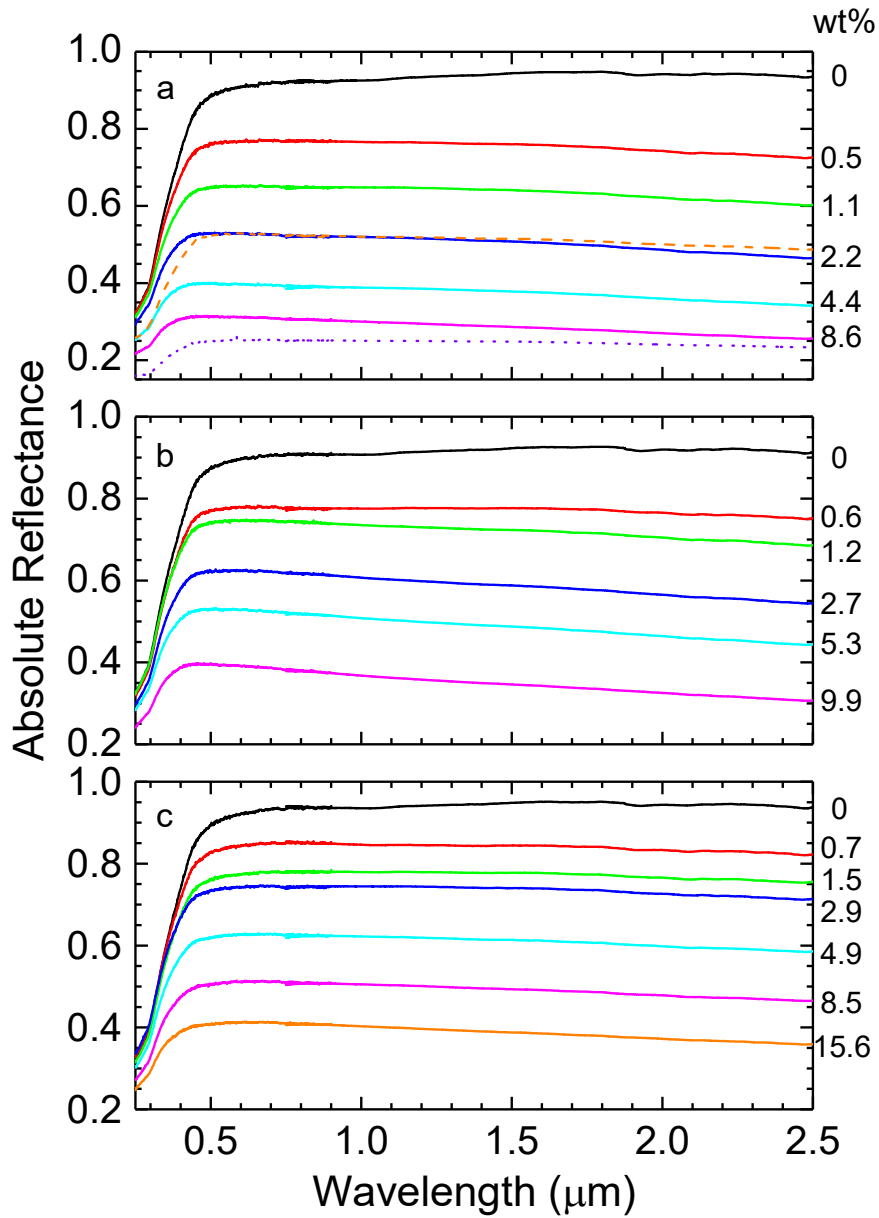

Figure 6. Absolute reflectance of Fo$_{99+}$ loose powder as it is mixed with dark material (< 45 μm): a) graphite, b) magnetite, and c) iron sulfide. The wt% of dark material is given to the right of each spectrum. Panel (a) also has two spectra (0.6 (dashed line) and 7.4 wt % graphite (dotted line)) from our experiment with 500 – 600 μm Fo$_{99+}$; all other spectra are using 45 – 125 μm Fo$_{99+}$.

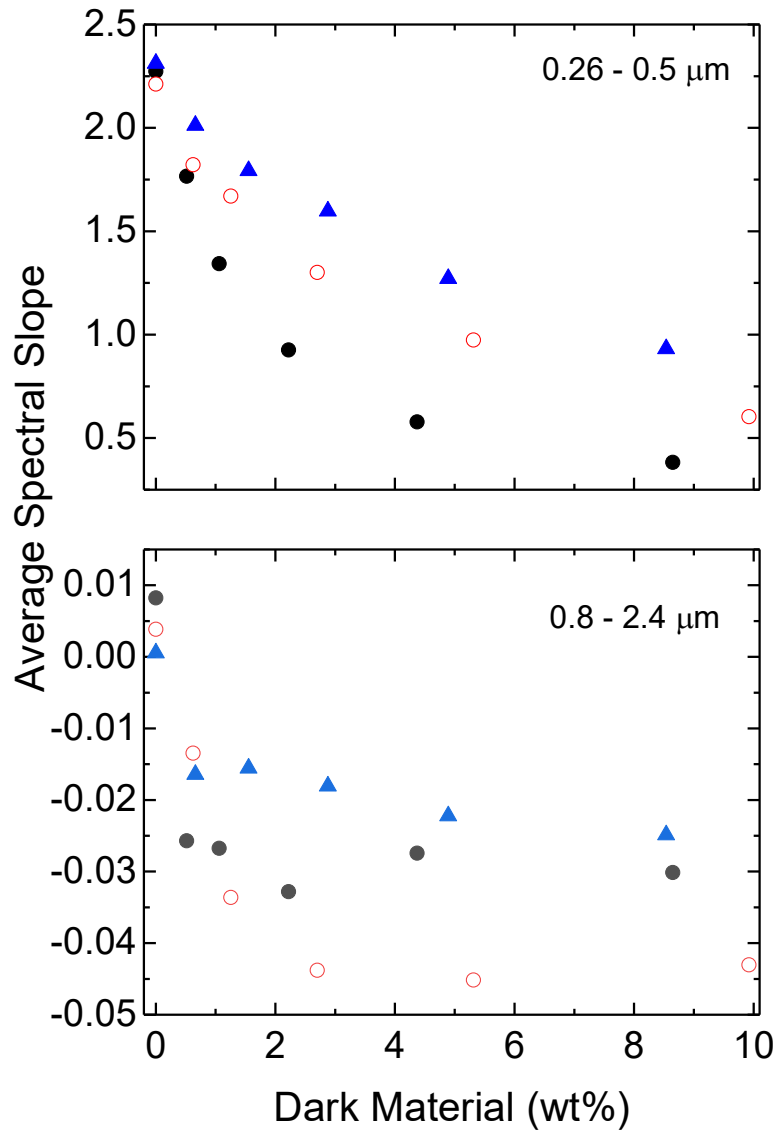

Figure 7. Changes in the average spectral slope in a $Fo_{99+}$ loose powder (45 – 125 μm) sample as it is mixed with the sieved (< 45 μm) dark material. The panels correspond to the average spectral slope between 0.26 and 0.5 μm (top) and 0.8 and 2.4 μm (bottom). In all panels, the symbols correspond to graphite (●), magnetite (○) and troilite (▲).

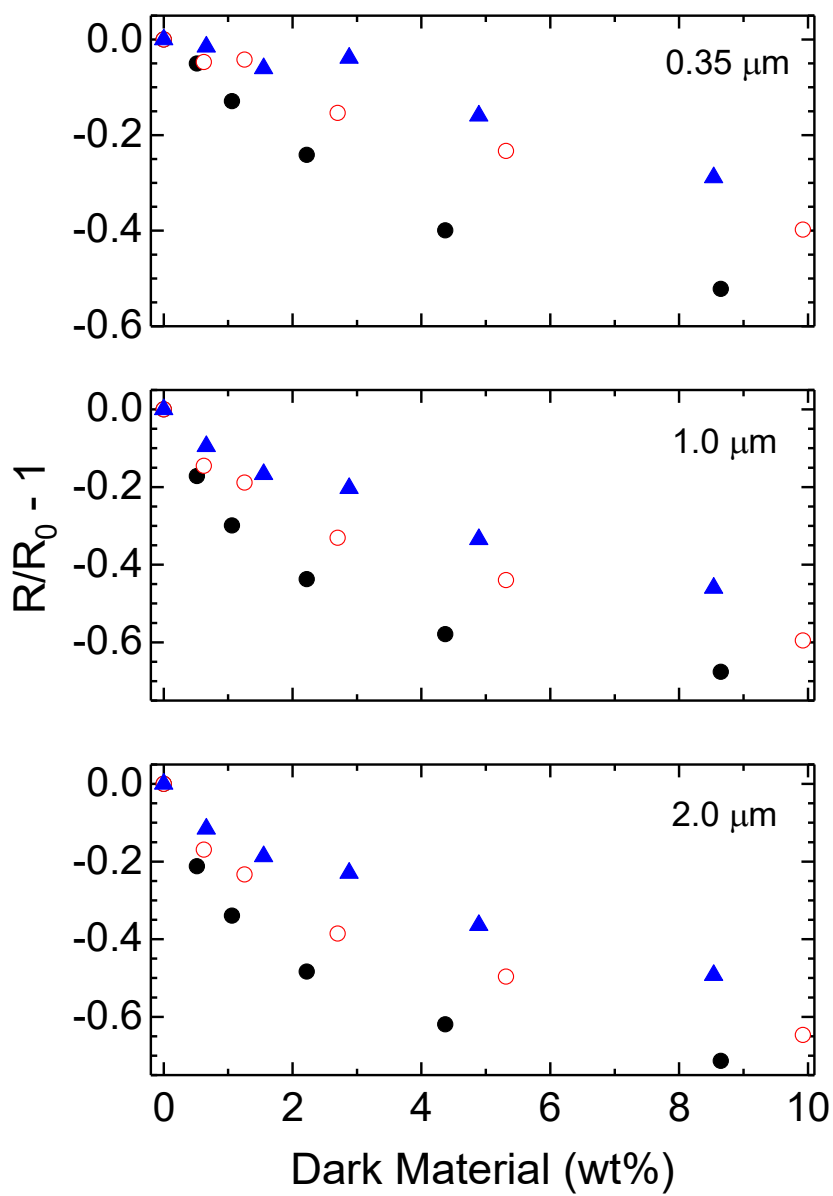

Figure 8. Changes in the normalized reflected intensity in a Fo$_{99+}$ loose powder (45 – 125 μm) sample as it is mixed with the sieved (< 45 μm) dark material. The panels (from top to bottom) correspond to the normalized reflected intensity at 0.35 μm, 1.0 μm, and 2.0 μm. In all panels, the symbols correspond to graphite (●), magnetite (○) and troilite (▲).

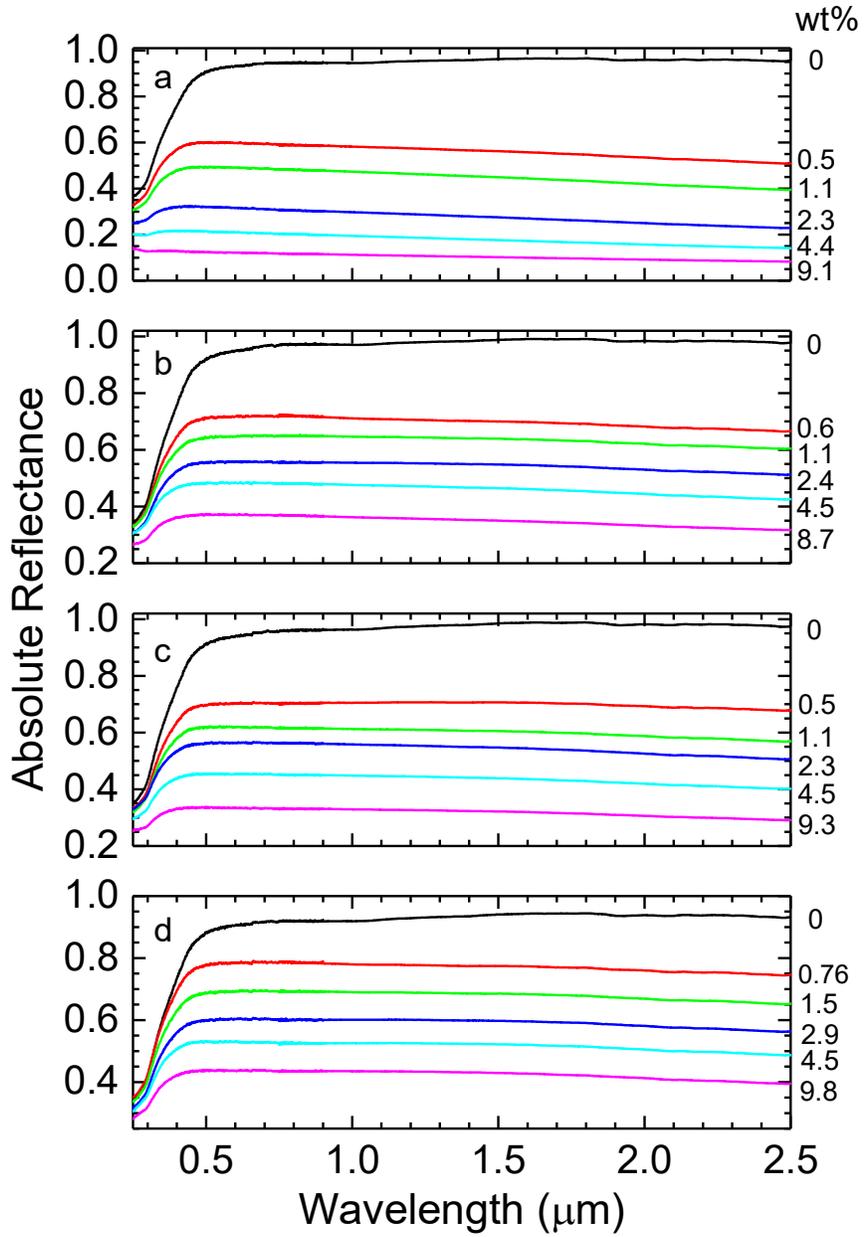

Figure 9. Absolute reflectance of Fo$_{99+}$ loose powder as it is mixed with graphite of the following grain sizes: a) 400 – 500 x 40 nm, b) 3 μm, c) 10 μm, and d) 25 μm. The wt% of graphite is given to the right of each spectrum.

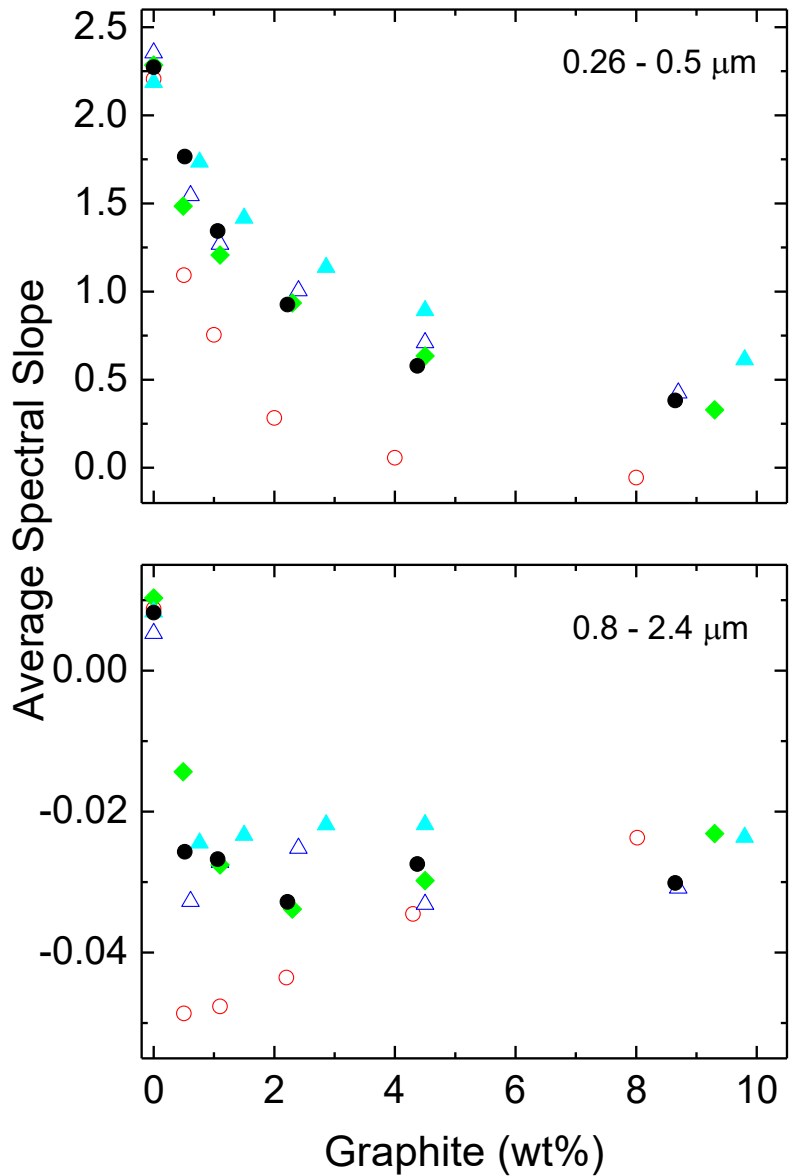

Figure 10. Changes in the average spectral slope in a Fo$_{99+}$ loose powder (45 – 125 μm) sample as it is mixed with graphite. The panels correspond to the average spectral slope between 0.26 and 0.5 μm (top) and 0.8 and 2.4 μm (bottom). In both panels, the symbols correspond to mixtures containing 400 – 500 x 40 nm (○), 3 μm (△), 10 μm (◆), 25 μm (▲) and < 45 μm (●) graphite grains.

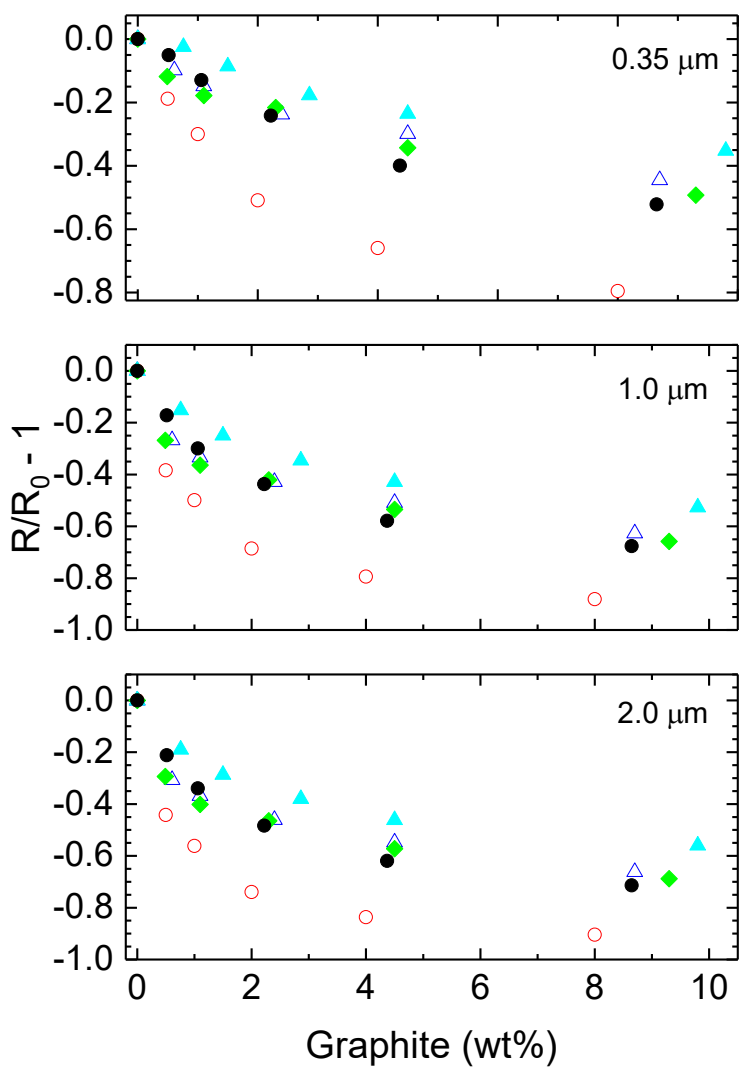

Figure 11. Changes in the normalized reflected intensity in a Fo$_{99+}$ loose powder (45 – 125 μm) sample as it is mixed with graphite. The panels (from top to bottom) correspond to the normalized reflected intensity at: 0.35 μm, 1.0 μm, and 2.0 μm. In all panels, the symbols correspond to mixtures containing 400 – 500 x 40 nm (○), 3 μm (△), 10 μm (◆), 25 μm (▲) and < 45 μm (●) graphite grains.

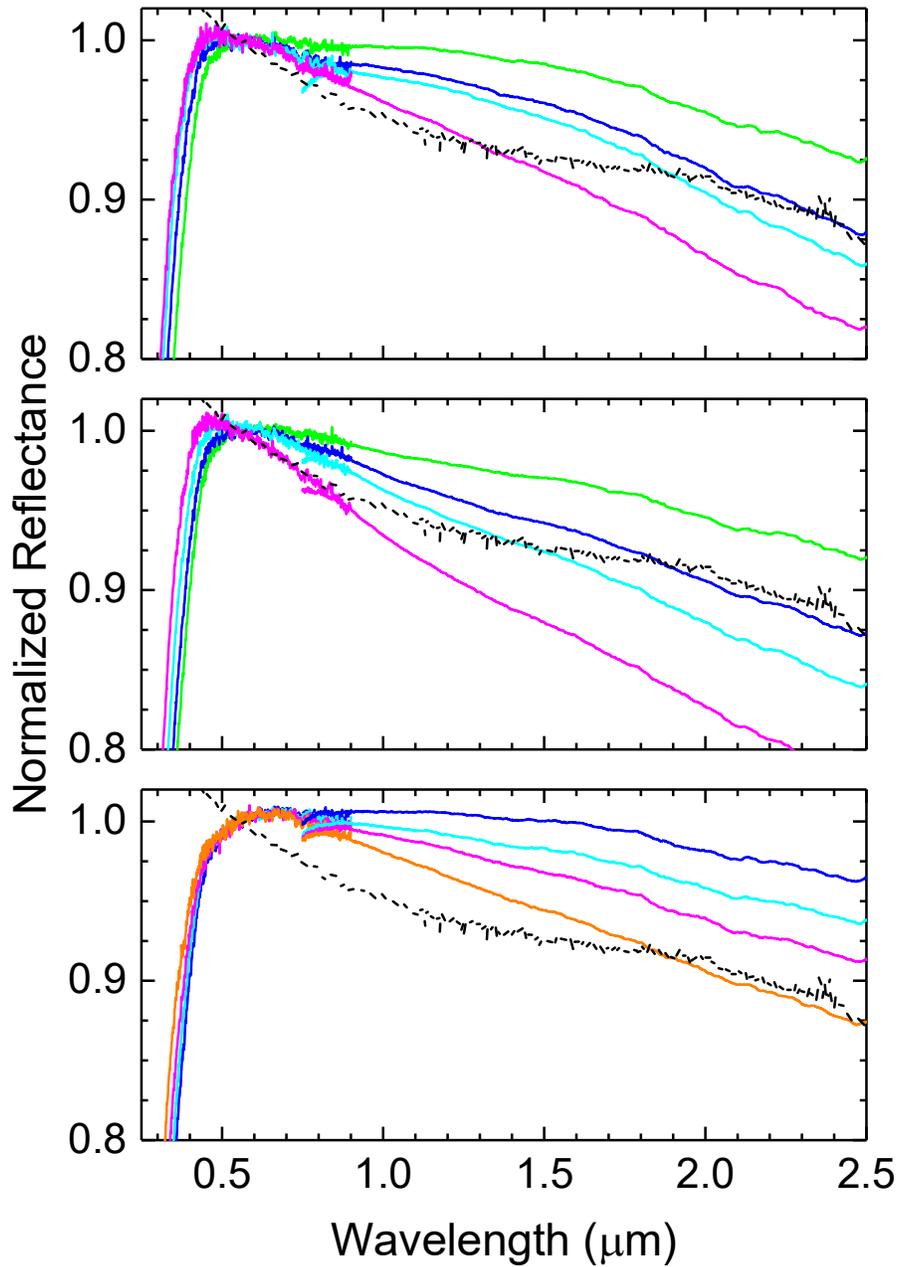

Figure 12. Reflectance spectra normalized at 0.55 μm of selected laboratory samples compared with a normalized spectrum of Bennu (dashed line) shown in Hamilton et al. 2019. Top (from top to bottom at 2.5 μm in wt% of graphite): 1.1, 2.2, 4.4, and 8.6. Middle (from top to bottom in wt% of $Fe_3O_4$): 1.2, 2.7, 5.3, and 9.9. Bottom (from top to bottom in wt% of FeS): 1.5, 2.9, 4.9, 8.5 and 15.6.